\documentstyle[preprint,prl,aps,epsf]{revtex}
\begin{document}
\tightenlines 
\title{Fourth Order Gradient Symplectic Integrator Methods\\ for 
Solving the Time-Dependent
Schr\"odinger Equation}
\author{ Siu A. Chin and Chia-Rong Chen}
\address{Center for Theoretical Physics, Department of Physics,\\ 
Texas A\&M University, 
College Station, TX 77843}
\maketitle
\begin{abstract}
We show that the method of splitting the operator
${\rm e}^{\epsilon(T+V)}$ to fourth order with purely positive coefficients
produces excellent algorithms for solving the 
time-dependent Schr\"odinger equation. These algorithms require knowing
the potential and the gradient of the potential. 
One 4th order algorithm only requires four Fast 
Fourier Transformations per iteration.
In a one dimensional scattering problem, the 4th order error
coefficients of these new algorithms are roughly 500 times smaller than
fourth order algorithms with negative coefficient, such as those based on
the traditional Ruth-Forest symplectic integrator. These algorithms can produce
converged results of conventional second or fourth order algorithms 
using time steps 5 to 10 times as large. Iterating these positive coefficient
algorithms to 6th order also produced better converged algorithms than iterating
the Ruth-Forest algorithm to 6th order or using Yoshida's 6th order algorithm A
directly.

\end{abstract}

\section {Introduction}

Understanding the dynamics of quantum evolution is of fundamental importance 
in all fields of physics and chemistry. Basic improvement in algorithms 
for solving the time-dependent Schro\"odinger equation can therefore 
impact many areas of basic research. Among numerical techniques 
developed for solving the time-dependent Schr\"odinger 
equation\cite{gold,feit,chev} (see T. N. Truong {\it et al.}\cite{truong} for  
earlier references), the method of split-operator\cite{feit}, or its
higher order variant, the method of symplectic 
integrator\cite{takah,gray,takah2},
has the advantage of being unitary, remain applicable in higher dimensions
and easily generalizable to higher order. The disadvantage is
that the time step size needed for convergence seemed to be small and many 
iterations are required for evolving system forward in time. 
In this work, We show that the method of factorizing the evolution operator
to fourth order with purely {\it positive} coefficients,
which have yielded a new class of gradient symplectic 
integrators for solving classical dynamical problems\cite{chin97,chindon}, 
also produces algorithms capable of solving the 
time-dependent Schr\"odinger equation with time steps 5 to 10 times
as large as before. 

The quantum state is evolved forward in time by the Schr\"odinger 
evolution operator
\begin{equation}
{\rm e}^{\epsilon H }={\rm e}^{\epsilon(T+V)},
\label{exph}
\end{equation}
where $\epsilon=-i\Delta t$, and 
$T=-{1\over2}\sum_i\nabla_i^2$, $V=V({\bf r}_i)$ are the kinetic
and potential energy operators respectively. (For clarity
of presentation, we will work in atomic units such that the kinetic energy 
operator has this standard form.)
In the split operator approach, the short-time evolution operator
 (\ref{exph}) is
factorized to second order in the product form
\begin{equation}
{\cal T}^{(2)}(\epsilon)\equiv
{\rm e}^{{1\over 2}\epsilon V}
{\rm e}^{\epsilon T}
{\rm e}^{{1\over 2}\epsilon V}={\rm e}^{\epsilon (T+V )+\epsilon^3 C+\cdots},
\label{second}
\end{equation}
where we have indicated the error term as $\epsilon^3 C$. Thus 
${\cal T}^{(2)}(\epsilon)$ evolves the system according to the 
Hamitonian $H^{(2)}=T+V+\epsilon^2C+\cdots$ 
which deviates from the original Hamiltonian by an error term second order 
in $\epsilon$.  
Since the kinetic energy operator is diagonal in momentum space, the 
split operator approach shuffles the wavefunction back and forth between 
real and Fourier space. 
(See detailed discussion by Takahashi and Ikeda\cite{takah}.)
Every occurrence of ${\rm e}^{\epsilon T}$ requires 
two Fast Fourier Transforms (FFTs), one
direct and one inverse. In this approach, the generalization
to higher dimension is straightforward, limited only by
the expense of higher dimensional Fourier transforms. Moreover, 
every factorization of the evolution operator 
${\rm e}^{\epsilon (T+V )}$ in the above form is unitary.

   One advantage of the split operator approach is that higher order 
algorithms can be constructed easily. For example, 
the evolution operator can be factorized to arbitrarily high order in the
form\cite{creutz,su,yoshid}
\begin{equation}
{\rm e}^{\epsilon (T+V )}=\prod_i
{\rm e}^{a_i\epsilon T}{\rm e}^{b_i\epsilon V},
\label{fact}
\end{equation}
with coefficients $\{a_i, b_i\}$ determined
by the required order of accuracy. This factorization process is 
identical to the derivation of symplectic algorithms for solving classical
dynamical problems\cite{yossym}. However, Suzuki\cite{nogo} has proved that, 
beyond second order, any factorization of the form (\ref{fact}) must produce 
some negative coefficients in the set $\{a_i, b_i\}$, corresponding to some
steps in which the system is evolved backward in time. While this is not 
detrimental in solving classical or quantum mechanical problems, it is
observed that in the classical case the resulting higher order 
symplectic algorithms 
converge only for very small ranges of $\Delta t$ and is far from 
optimal\cite{chindon}. As we will show below, the same is true for
quantum algorithms. In this work, we show that insisting on factorizing 
the the Schr\"odinger evolution operator to 4th order with purely 
{\it positive} time steps yielded algorithms with
excellent convergent properties at large time steps.

\section {Fourth Order Operator Splittings}

An example of 4th order splitting with negative coefficient is the 
Ruth-Forest\cite{forest} scheme,
\begin{equation}
{\cal T}_{FR}^{(4)}(\epsilon)={\cal T}^{(2)}(\widetilde\epsilon)
{\cal T}^{(2)}(-s\widetilde\epsilon)
{\cal T}^{(2)}(\widetilde\epsilon)
\label{rf}
\end{equation}
where
$s=2^{1/3}$ is chosen to cancel the $\epsilon^3C$ error term in 
${\cal T}^{(2)}$ and $\widetilde\epsilon=\epsilon/(2-s)$ rescales
the sum of forward-backward-forward time steps back to $\epsilon$.
This factorization scheme has been independently derived many times
in the context of symplectic integrators\cite{camp,candy}.
The above derivation was first
published by by Creutz and Gocksch\cite{creutz} in 1989. Suzuki\cite{su}
and Yoshida\cite{yoshid} independent published the same 
constructions in 1990. 
Identical construction can be applied to generate
a $(n+2)$th order algorithm
${\cal T}^{(n+2)}$ from a triplet products of ${\cal T}^{(n)}$'s,
\begin{equation}
{\cal T}^{(n+2)}(\epsilon)={\cal T}^{(n)}(\widetilde\epsilon)
{\cal T}^{(n)}(-s\widetilde\epsilon)
{\cal T}^{(n)}(\widetilde\epsilon)
\label{higher}
\end{equation}
with $s=2\,^{1/(n+1)}$. The Ruth-Forest (RF) algorithm requires 6 FFTs. 
The alternative algorithm with operators $V$ and $T$ interchanged is 
also possible, but would have required 8 FFTs per iteration.  

Recently, Suzuki\cite{suzukiab} and Chin\cite{chin97} have 
derive a number of 4th order splitting schemes with only positive 
coefficients. In order to circumvent Suzuki's ``no positive coefficient" 
proof, these factorizaztions
require the use of an additional operator $[V,[T,V]]=\sum_i|\nabla_iV|^2$, 
which means that these new algorithms require knowing the gradient of
the potential. The two schemes derived by both Susuki and Chin, using 
different methods, are:    
\begin{equation}
{\cal T}_{A}^{(4)}\equiv 
  {\rm e}^{\epsilon {1\over 6} V}
  {\rm e}^{\epsilon {1\over 2} T} 
  {\rm e}^{\epsilon {2\over 3} \widetilde V } 
  {\rm e}^{\epsilon {1\over 2} T} 
  {\rm e}^{\epsilon {1\over 6} V},
\label{foura}
\end{equation}
with $\widetilde V$ given by
\begin{equation}
\widetilde V=V+{1\over 48}\epsilon^2[V,[T,V]],
\label{superv}
\end{equation}
and
\begin{equation}
{\cal T}_{B}^{(4)}\equiv 
  {\rm e}^{\epsilon {1\over 2}(1-{1\over{\sqrt 3}}) T}
  {\rm e}^{\epsilon {1\over 2} \bar V } 
  {\rm e}^{\epsilon {1\over{\sqrt 3}}   T}
  {\rm e}^{\epsilon {1\over 2} \bar V } 
  {\rm e}^{\epsilon {1\over 2}(1-{1\over{\sqrt 3}}) T},
\label{fourb}
\end{equation}
with $\bar V$ given by
\begin{equation}
\bar V=V+{1\over 24}(2-\sqrt 3)\epsilon^2[V,[T,V]]. 
\label{duperv}
\end{equation}
Note that scheme A, remarkably, only requires 4 FFTs. Chin's
splitting scheme C,
\begin{equation}
{\cal T}_{C}^{(4)}\equiv 
  {\rm e}^{\epsilon {1\over 6} T} 
  {\rm e}^{\epsilon {3\over 8} V}
  {\rm e}^{\epsilon {1\over 3} T} 
  {\rm e}^{\epsilon {1\over 4} \widetilde V}
  {\rm e}^{\epsilon {1\over 3} T} 
  {\rm e}^{\epsilon {3\over 8} V}
  {\rm e}^{\epsilon {1\over 6} T},
\label{chinc}
\end{equation}
which minimizes the appearance of $V$ for the derivation of
symplectic algorithms, has 4 $T$ operators, corresponding to
8 FFTs. This is undesirable in the current context. It is however
easy to derive an alternate 4th order scheme with only 3 $T$ operators by
splitting the operator product at midpoint and concatenate the ends
together to yield 
\begin{equation}
{\cal T}_{D}^{(4)}\equiv 
  {\rm e}^{\epsilon {1\over 8} \widetilde V}
  {\rm e}^{\epsilon {1\over 3} T} 
  {\rm e}^{\epsilon {3\over 8} V}
  {\rm e}^{\epsilon {1\over 3} T}
  {\rm e}^{\epsilon {3\over 8} V}
  {\rm e}^{\epsilon {1\over 3} T} 
  {\rm e}^{\epsilon {1\over 8} \widetilde V}.
\label{chind}
\end{equation}
This ``split and splice" operation only works on scheme C because this
scheme was originally derived by symmetrizing the splitted product. 
Scheme D is just the other way of symmetrizing the same product. 
These two algorithms gave identical results in the scattering problem solved
below. Obviously then, algorithm 4D is preferable with two fewer FFTs.

\section {One Dimensional Scattering}

To gauge the effectiveness of these new algorithms, we test them
on a one dimensional scattering problem, 
where a Gaussian wave pocket
\begin{equation}
\psi_0(x) ={1\over{ (2\pi\sigma^2)^{1/4}}}
\exp\left [ ik_0x-{{(x-x_0)^2}\over{4\sigma^2}}\right ],
\label{gwave}
\end{equation}
is impinged on a smooth sech-square potential. 
The Hamiltonian is given by 
\begin{equation}
H=-{1\over 2}{{d^2}\over{dx^2}}+V_0\,{\rm sech}^2(x)
\label{ham}
\end{equation}
This choice of the potential is dictated by the fact that its
transmission coefficient is known analytically\cite{takah2}
\begin{equation}
T={ 1\over{	1+\cosh^2(\pi\sqrt{2V_0-1/4})/\sinh^2(\pi k_0) }}.
\label{tc}
\end{equation}
We choose $V_0=48.2$ so that when the initial energy $E_0={1 \over 2}k_0^2$
is equal to $V_0$, the transmission coefficient is 0.520001, which is
practically 0.52 for our purpose.

To compute the transmission coefficient, we evolve the Gaussian 
wave pocket initially sufficiently far from the barrier and then
integrate the transmitted wave pocket after a time of $t_{max}=20$,
when the latter is well separated from the reflected wave . We use 
$2^{12}=4096$ grid points over a length of 600, yielding 
a discretization spacing of $\Delta x \approx 0.15$. Using more  
grid points than this has no measurable impact on the final results.
We found that in order
to reproduce the analytical transmission coefficient, it is necessary to
use a very flat Gaussian incident wave pocket. We therefore take 
$\sigma=20$ and place the wave pocket initially at $x_0=-80$.

Fig. 1 shows the resulting transmission coeffcient for various algorithms
as a function of the time step size $\Delta t$ at an incident energy of
$E_0=V_0$. Even with such a flat Gaussian incident wave pocket, at the 
smallest time step for the best algorithm, the transmission coefficient 
converges only to $T_0=0.519905$ . While this value is still slightly 
below the exact value due to a finite sized Guassian 
wave pocket, it is a perfectly acceptable benchmark to compare all 
algorithms with identical starting conditions.

The second order results (\ref{second}), denoted by asterisks, can be
accurately fitted by $T_0-0.36\Delta t^2$ for $\Delta t<0.1$, demonstrating its 
quadratic convergence. The results of the Ruth-Forest scheme (\ref{rf}),
can also be well fitted by $T_0-74\Delta t^4$ over the same range as shown, 
verifying it quartic convergence. However, it is clearly obvious that
the range of convergence of the RF algorithm is not substantially 
greater than that of the second order algorithm, perhaps at most a factor 
of three greater.
In comparison, the four 4th order algorithms with positive splitting 
coefficients are distinctly superior. Whereas 
the fourth order error coefficient of the Ruth-Forest algorithm
is 74, the corresponding coefficients
for algorithms 4A, 4B, 4C and 4D are respectively, -1.07, -0.38, 0.14 and 0.14
respectively. Algorithm 4C and 4D yielded identical results. Algorithm 
4D's error cofficient is more than 500 times smaller than that of RF, 
and can achieve the same accuracy by using step sizes nearly 5 times as large. 
The comparison with second order results is even more favorable; 
the step size can be 10-15 times as large. To compare the computational 
effort involve, we timed each algorithm for 160 iterations on a Pentium II
450 MH processor using a Fortran compiler. The second order algorithm took
$T_2=5.33$s. Relative to this time, the time required by algorithms RF, 4A, 
4B, 4C, 4D are respectively, 2.98$T_2$, 2.22$T_2$, 3.37$T_2$, 3.97$T_2$,
and 3.26$T_2$ respectively, which roughly scale with the number of FFTs 
used in each algorithm. Algorithm 4A is specially notable in that it is 
roughly 1/3 faster than RF but converges at time steps nearly 10 times as large. 
We have used algorithm 4A at time step size $\Delta t=0.1$ to compute the
transmission coefficient as a function of the incident energy.
Over the range of $E_0/V_0=0.8$ to 1.2, where the transmission coefficient
goes from 0.0016 to 0.9974, the results are in agreement 
with the exact value (\ref{tc}) to at least three decimal places.

At present, no 6th order factorization with positive coefficients are known.
However, one can use the triplet construction (\ref{higher}) to build a 
6th order algorithm by iterating on three 4th order algorithms. Fig. 2 shows the
resulting convergence curves for various 6th order algorithms. The solid triangles 
corresponds to iterating on the RF algorithm to 6th order (RF6). 
There is no visible improvement in the convergence range. 
This algorithm requires 18 FFTs. The asterisks are 
Yoshida's\cite{yoshid} 6th order algorithm A (Y6A) , which is a product of 
7 second order algorithms (\ref{second}) some with negative coefficients, 
requiring 14 FFTs. 
Its convergence range is about twice that of the RF6 algorithm. The hollow 
diamonds, hollow circles, and solid circles are 6th order results based on 
algorithms 4A, 4B, and 4D respectively, and will be referred to as due to
algorithms 6A, 6B, and 6D respectively.
Note that algorithm 6A only requires 12 FFTs. 
By fitting a polynomial of orders 6 to 12 in $\Delta t$, we extracted the 6th order error
coefficients for each algorithms. For algorithms RF6, Y6A, 6A, 6B and 6D, 
the error coefficients are -7675, -171, -17.42,
-6.887, and 5.819 respectively. The new, gradient algorithms are orders of magnitude
better than previous 6th order algorithms. Algorithm 6B's results are so flat 
that they can be fitted by a polynomial in $(\Delta t)^8$ alone. For comparsion, 
we have also replotted the 4th order results due algorithm 4D as a dashed line. 
Algorithms RF6 and Y6A are not even better than 4D.  
Since all these 6th order algorithms, with the exception of Y6A, are just the 
product of three 4th order algorithms, their running time simply triple that 
of their respective 4th order algorithm's running time. Y6A's time is obviously
7$T_2$, which is faster than all other algorithms except 6A.    

\section {Conclusions}

In this work, we have demonstrated that 4th order split operator algorithms, with
no negative intermediate time steps, are superior to existing second order or 
fourth order algorithms for solving the time-dependent Schr\"odinger equation.
It is straighforward to generalize these algorithms to higher dimension by 
using higher dimensional FFTs. These new algorithms require
calculating the gradient of the potential, but converge at much
large time step sizes. They should be useful for rapid simulation of large 
quantum systems with relatively simple potentials.

Our comparison of 6th order algorithms suggests that higher order 
algorithms with intermediate negative time steps are far from optimal.
Algorithms RF6 and Y6A, which uses more negative time steps, are 
inferior to algorithm 6A, 6B, or 6D. This is the same conclusion drawn 
recently when higher order symplectic algorithms are 
compared in solving the Kepler problem\cite{chindon}.
This will impact current interests in implementing higher 
order symplectic
algorithms to study quantum dynamics\cite{takah,gray,takah2}.  
  
This work suggests that the continual search for 
purely positive coefficients factorization schemes may produce better 
converged algorithms for solving both classical and quantum dynamical problems.
Currently, there are no known 6th order splitting schemes with purely positive 
coefficients.

\acknowledgements
This work was supported, in part, by the National Science Foundation
grants No. PHY-9870054 to SAC.


\ifpreprintsty\newpage\fi
\begin{figure}
\noindent
\vglue 0.2truein
\hbox{
\vbox{\hsize=7truein
\epsfxsize=6truein
\leftline{\epsffile{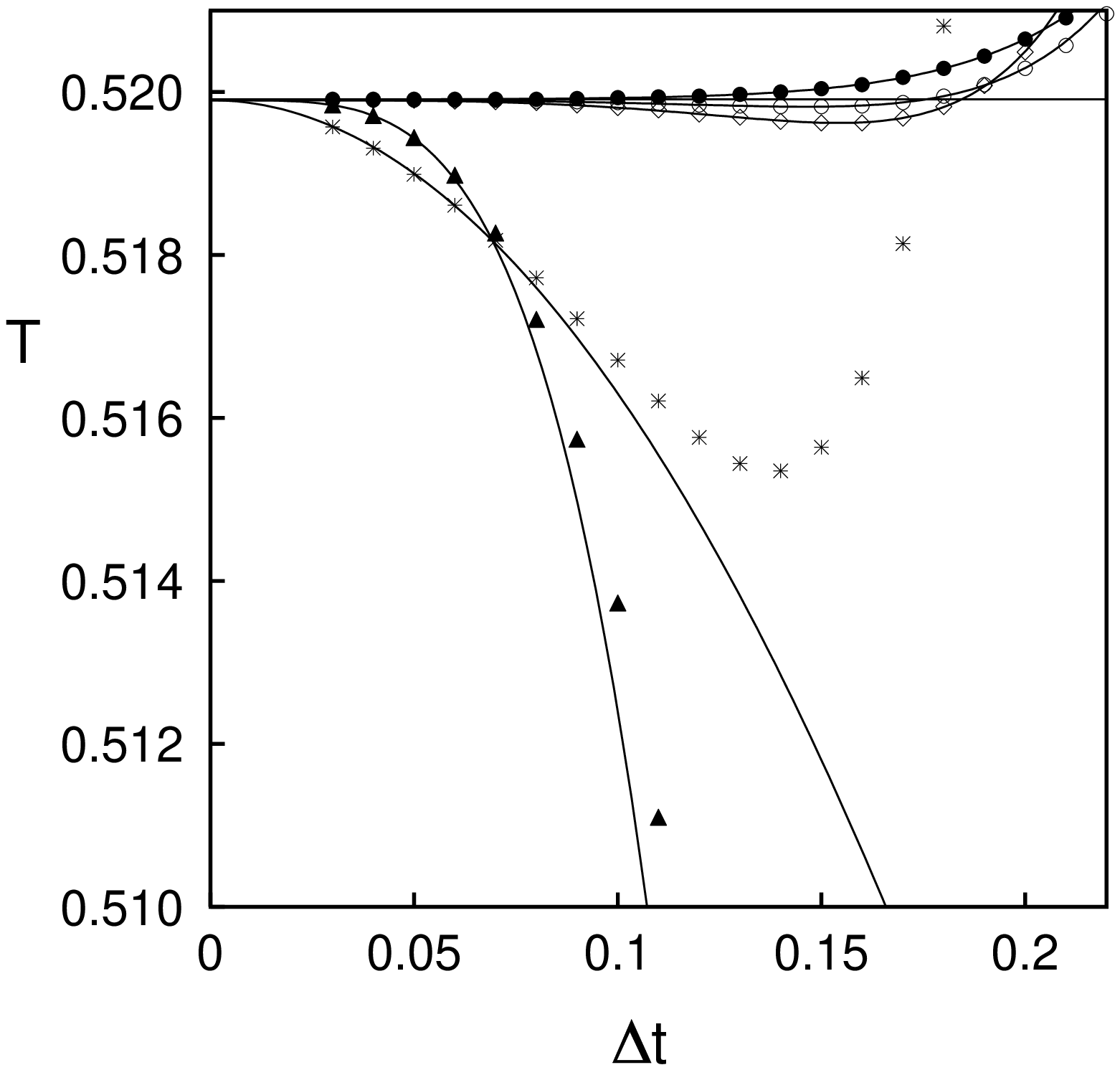}}
}
}
\vglue 0.3truein
\caption{The transmission coefficient $T$ as a function of
time step size for various split operator algorithms. The asterisks
are second order results, (\ref{second}). The solid triangles are
4th order results corresponding to the Ruth-Forest splitting scheme 
with negative coefficients, (\ref{rf}). The hollow diamonds and circles
are results of algorithm 4A, (\ref{foura}) and 4B, (\ref{fourb})
respectively. The filled circles are identical results produced by
algorithms 4C, (\ref{chinc}), and 4D, (\ref{chind}). 
The lines are fitted lines to extract the leading
error coefficients. See text for further details.
}
\label{fone}
\end{figure}
\ifpreprintsty\newpage\fi
\begin{figure}
\noindent
\vglue 0.2truein
\hbox{
\vbox{\hsize=7truein
\epsfxsize=6truein
\leftline{\epsffile{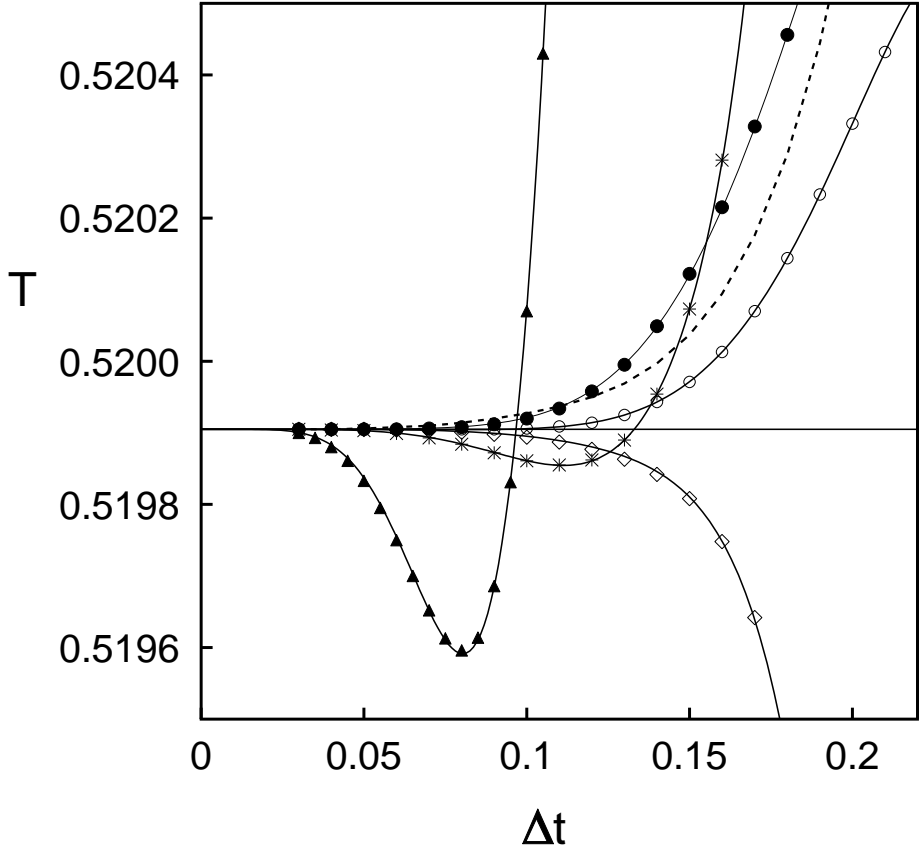}}
}
}
\vglue 0.3truein
\caption{The transmission coefficient $T$ as a function of
time step size for various iterated 6th order algorithms.
The sold triangle are results of a 6th order algorithm based on the
4th order Ruth-Forest algorithm. The asterisks corresponds to Yoshida 6th order
algorithm A. The hollow diamonds, hollow circles, and solid circles, are 6th order
algorithm results based on iterating the 4th order algorithm 4A, 4B, and 4D respectively.
See text for further details. The solid lines are fitted polynomials in powers
$\Delta t$ beginning with powers of 6 up to 12. For comparison, the dash line 
corresponds to the best of the 4th order results, due to algorithm 4D.  
 }
\label{ftwo}
\end{figure}
\end{document}